%% file: main.tex
\documentclass[a4paper]{article}

\usepackage{INTERSPEECH2022}

\usepackage{array}

\usepackage{enumitem}
\usepackage{slantsc}
\usepackage{multirow}
\usepackage{dsfont}
\usepackage{color}
\usepackage{arydshln}
\usepackage{hyperref}
\usepackage{booktabs}
\usepackage{cite}
\usepackage[tracking=true]{microtype}

\usepackage{etoolbox}
\title{\vspace{-2.5ex} Audio Similarity is Unreliable as a Proxy for Audio Quality \vspace{-1ex}}
\name{Pranay Manocha$^1$, Zeyu Jin$^2$, Adam Finkelstein$^1$}
\address{
  $^1$Department of Computer Science, Princeton University, USA\\
  $^2$Adobe Research, USA}
\email{{\{pmanocha,af\}}@cs.princeton.edu, zejin@adobe.com}

\DeclareUnicodeCharacter{2212}{-}
\input{macros.tex}

\begin{document}


\maketitle
\input{LaTeX/texts/00-abstract-test}

\noindent\textbf{Index Terms}: audio quality, speech quality, similarity metrics, perceptual metric, speech enhancement 

\input{LaTeX/texts/01-intro-test}
\input{LaTeX/texts/03-Similarity-and-Distance}

\input{LaTeX/texts/04-results}
\input{LaTeX/texts/06-conclusions}


\clearpage
\bibliographystyle{IEEEtran}

\bibliography{mybib}


\end{document}

%% file: macros.tex
\newcommand{\ignorethis } [1] {}

\newcommand{\sectnum    } [1] {\ref{#1}}

\newcommand{\fignum     } [1] {\ref{#1}}

\newcommand{\sect       } [1] {Section~\sectnum{#1}}

\newcommand{\fig        } [1] {Figure~\fignum{#1}}

\newcommand{\etal       }     {{et~al.}}

\newcommand{\eg         }     {{e.g.}}

\newcommand{\ie         }     {{i.e.}}

\newcommand{\Reals      }     {{\textrm{I\kern-0.18em R}}}

\renewcommand{\vec      } [1] {{\text{\boldmath $\mathbit{#1}$}}}

\newcommand{\change     } [1] {\mbox{{\footnotesize $\Delta$} \kern-3pt}#1}

\newlength{\w}

\newcommand\narrowstyle{\SetTracking{encoding=*}{-50}\lsstyle}

\newcommand{\acronym}[1] {{\uppercase{#1}}}

\newcommand{\PESQ}   {\acronym{Pesq}}
\newcommand{\VISQOL} {\acronym{Visqol}}
\newcommand{\POLQA}  {\acronym{Polqa}}
\newcommand{\STOI}   {\acronym{Stoi}}

\newcommand{\DPAM}   {\acronym{Dpam}}
\newcommand{\CDPAM}   {\acronym{Cdpam}}
\newcommand{\NORESQA}   {\acronym{Noresqa}}
\newcommand{\DNSMOS}   {\acronym{Dnsmos}}
\newcommand{\NISQA}   {\acronym{Nisqa}}
\newcommand{\SQAPP}   {\acronym{Sqapp}}

\newcommand {\largefontsize}[1]{\fontsize{16}{36pt}\selectfont{#1}}
\newcommand {\largefontsizelower}[1]{\fontsize{12}{36pt}\selectfont{#1}}

\newcolumntype{$}{>{\global\let\currentrowstyle\relax}}

\newcolumntype{^}{>{\currentrowstyle}}

\newcommand{\rowstyle}[1]{\gdef\currentrowstyle{#1}%
  #1\ignorespaces
}

%% file: LaTeX/texts/00-abstract-test.tex
\begin{abstract}
  
  
  Many audio processing tasks require perceptual assessment. 
  However, the time and expense of obtaining ``gold standard'' human judgments limit the availability of such data. 
  Most applications incorporate full reference or other similarity-based metrics (\eg\ PESQ) that depend on a clean reference.
  Researchers have relied on such metrics to evaluate and compare various proposed methods, often concluding that small, measured differences imply one is more effective than another.
  %
  %
  This paper demonstrates several practical scenarios where similarity metrics fail to agree with human perception, because they: (1)~vary with clean references; (2)~rely on attributes that humans factor out when considering quality, and (3)~are sensitive to imperceptible signal level differences. 
  In those scenarios, we show that no-reference metrics do not suffer from such shortcomings and correlate better with human perception.
  %
  %
  %
  %
  We conclude therefore that similarity serves as an unreliable proxy for audio quality.

\end{abstract}

%% file: LaTeX/texts/01-intro-test.tex
\section{Introduction}

Speech quality assessment (SQA) plays a critical role in a range of audio and speech applications, including telephony, VoIP, hearing aids, automatic speech recognition, and speech enhancement. The ``gold standard'' for SQA involves human listening tests. However, these subjective evaluations are time consuming and expensive, as they require a human in the loop and often need to be repeated many times for every recording.
Thus automatic (objective) SQA methods are often more practical, and roughly fall into two categories -- whether or not they rely on a reference recording. 

\emph{Full-reference} metrics are also known as \emph{intrusive} or \emph{similarity} metrics (\eg\ \PESQ~\cite{rix2001perceptual}, \POLQA~\cite{beerends2013perceptual}, \VISQOL~\cite{hines2015visqol}, \DPAM~\cite{manocha2020differentiable}, \CDPAM~\cite{manocha2021cdpam} and others~\cite{patton2016automos,fu2019metricgan,yu2021metricnet}).
They require a clean reference to which a corrupted signal can be compared as the basis for a quality rating. On the other hand, \emph{no-reference} metrics, also called \emph{non-intrusive}~\cite{loizou2011speech} metrics, (\eg\ \DNSMOS~\cite{reddy2020dnsmos}, \NISQA~\cite{mittag2021nisqa}, \SQAPP~\cite{Manocha:2022:SNS} and others~\cite{fu2018quality,andersen2018nonintrusive,lo2019mosnet,gamper2019intrusive,zhang2021end}) are designed to output a rating on an absolute scale, without access to a clean reference. 
Researchers commonly rely on full-reference metrics as a proxy for audio quality, because they were introduced earlier~-- consider, \eg\ SNR.
One of the most impactful is \PESQ~\cite{rix2001perceptual}, introduced decades ago for telephony and still used today for enhancement~\cite{su2020hifi,manocha2020differentiable,manocha2021cdpam,manocha2021noresqa,defossez2020real,zezario2021deep}, vocoders~\cite{cernak2005evaluation}, and transmission codecs~\cite{beerends2004measurement,paglierani2007uncertainty}. \PESQ\ correlates well with subjective listening tests~\cite{reddy2020dnsmos,fu2018end}, and
\PESQ\ labels have also been leveraged for training \emph{differentiable} quality metrics~\cite{fu2018quality,fu2019metricgan,xu2021deep}. 
%

%

%

\input{LaTeX/tables/scenario1}

Researchers rely on objective metrics like \PESQ\ for comparing the effectiveness of various methods, sometimes reporting small gains ($\sim$0.1 \PESQ).
However, \PESQ\ has acknowledged shortcomings~\cite{manjunath2009limitations,hines2013robustness}, and may not be reliable to detect subtle differences~\cite{manocha2020differentiable}.
Researchers have explored ways to ameliorate such shortcomings and improve robustness of such objective metrics~\cite{beerends2013perceptual,manocha2020differentiable}.
However, this paper describes experiments suggesting that the inherent problem may be the overall formulation relying on a clean reference.
For example, \fig{scenario1} sketches a scenario where two different ``clean'' reference recordings are used to evaluate the quality of a test recording. An effective metric should report the same quality for the test, as long as the reference is ``clean'' -- but similarity metrics naturally report different quality measures in this scenario, because the recording setup of the test happens to be acoustically closer to one of them.


This paper investigates such limitations of similarity metrics, with the goal of informing future SQA research.
%
%
We describe several scenarios wherein we empirically evaluate seven similarity metrics (L1, L2, Multi-resolution STFT, \PESQ, \VISQOL, \DPAM\ and \CDPAM), and show that these contrast with subjective quality judgments.
Overall, we find that: (1)~similarity metrics fail to capture the multi-modality of audio quality relative to ``clean'' recordings made in different environments; (2)~mismatched training and evaluation datasets cause objective ratings to differ from subjective ratings; and (3)~similarity metrics emphasize imperceptible differences, and therefore fail for recordings that sound perceptually indistinguishable but is sampled from a different distribution.
%
%
Moreover, we also show specific examples (\eg\ high-pass filtering or comparing cross-perturbations) where similarity metrics contrast with subjective judgments. 
%
%
We also evaluate four no-reference metrics (\DNSMOS, \NISQA, \SQAPP\ and \NORESQA) on those scenarios, and show that they do not suffer from the same shortcomings. 
%
%
These findings illustrate some weaknesses of similarity metrics (\eg\ \PESQ), and help inform researchers on how best to use them across various tasks. 
%
%
%
Listening examples are available here:
\\{\footnotesize\tt\narrowstyle \href{https://pixl.cs.princeton.edu/pubs/Manocha_2022_ASI/}{https://pixl.cs.princeton.edu/pubs/Manocha\_2022\_ASI/}}

%% file: LaTeX/tables/scenario1.tex
\begin{figure}[t!]
\centering
\setlength{\tabcolsep}{4pt}
\includegraphics[width=\columnwidth]{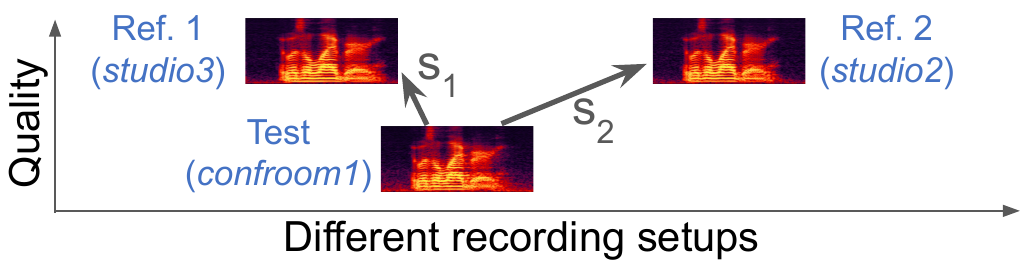}
\vspace{-0.25in}

\caption{\textbf{Scenario 1}: Ref. 1 and Ref. 2 are recordings from different environments from the DDS dataset~\cite{li2021dds} (\textit{studio2} and \textit{studio3}), of equal quality but with different acoustic settings. A random recording from a different setting (\textit{confroom1}), may have different similarities ($s_1$ and $s_2$) with the two references, even though both have equal quality. (Adapted from ~\cite{Manocha:2022:SNS})}
\vspace{0\baselineskip}
\label{scenario1}
\end{figure}

%% file: LaTeX/texts/03-Similarity-and-Distance.tex
\section{When similarity does not imply quality}

Measurement of similarity typically involves a pair of audio signals, a test signal and a reference. Typically, the test signal is created in relation to the reference, meaning that they share the same content and are synchronized in phase. Quality measures, such as MOS, may not involve a reference that is in parallel with the test signal. This difference informs us of two potential issues in using similarity as proxy for quality: (1)~different references, although sharing the same quality, may result in different similarities when compared with the same test signal, and (2)~different test signals, although having the same quality rating may have significantly different similarities to a reference. As a particular but common case of (2), two signals that are perceptually indistinguishable may have different similarities with a reference due to imperceptible factors such as slight difference in frequency energy, inaudible noises and phase differences. We examine the two cases and the special case in Sections~\ref{subsec2.1},~\ref{subsec2.2} and~\ref{subsec2.3} respectively. 


%

In the first scenario, we use audio samples recorded in two different environments (\textit{studio2} and \textit{studio3} in DDS~\cite{li2021dds} dataset) as references to evaluate similarity with test examples recorded in a noisy (\textit{confroom1}) environment. While the audio quality of the references are close based on MOS ratings, the similarity metrics differ significantly.
The second scenario is a widely used case of evaluating and comparing quality of trained models on a common evaluation dataset (\eg\ VCTK for speech enhancement). Here we show that a lower quality test recording may be rated as more `similar' (or higher quality) to a given clean reference, than a higher quality test recording. 
%
%
Finally, we investigate how the similarity metrics vary when we acoustically process one dataset to sound like another. We observe that similarity metrics still show a wide variation even when the recordings become perceptually indistinguishable, suggesting that these metrics emphasize imperceptible differences at the signal level.

To show the generality of the issues, we utilize a wide variety of similarity metrics - including conventional methods like L1, L2 and Multi-resolution STFT, as well as perceptually driven metrics like \PESQ, \VISQOL, \DPAM\ and \CDPAM. In the analysis, we also employ various no-reference metrics like \SQAPP, \NISQA\, \DNSMOS\, as well as the non-matching no-reference framework \NORESQA. For each scenario, we compute these objective metrics, as well as subjective listening scores. Lets look at each scenario in detail.

\input{LaTeX/tables/scenario2}

\input{LaTeX/tables/dds_datset_environment}

\subsection{Two reference with one test recording}
\label{subsec2.1}

Audio quality is multi-modal, \ie\ two different signals can share the same quality rating but sound perceptually different. However, similarities can vary based on different references even when these references share similar quality. 
See~\fig{scenario1}.

%
%

In this scenario, we use the newly released DDS dataset~\cite{li2021dds} which provides aligned parallel recordings of studio-quality speech and their lower-quality versions recorded in 9 environment with 3 devices at six microphone positions (totaling 2000 hours of speech samples). The fact that these samples are produced from and synchronized with clean audio samples gives us the ability to apply various similarity metrics. We choose each device and microphone position and compute all similarity metrics between lower-quality samples and the clean reference. To get the ground truth quality rating, we conduct an MOS test on recordings from each environment. Then we choose a pair of environments that are close in quality (\textit{studio2} and \textit{studio3}) as reference groups and a lower-quality (\textit{confroom1}) environment as test group. Finally, we compute similarities between test group and reference groups as well as no-reference quality metrics as comparison. The results are summarized in Section~\ref{sec4.1}.

%
%
%
%

\subsection{One reference with two test recordings}
\label{subsec2.2}

\fig{scenario2} illustrates a scenario in which we hypthesize that similarity and quality can have negative correlation -- comparing two signals output from different processes with the a single reference.
Our experiment involves \emph{speech enhancement} (SE),
whose goal is to improve the perceptual quality of speech signals by removing background noise and other perturbations. 
Because subjective evaluations are not comparable across different experimental settings~\cite{cooper2022generalization}, researchers rely on objective metrics (\eg\ \PESQ\ for quality and \STOI~\cite{taal2010short} for intelligibility) on a common dataset (\eg\ VCTK~\cite{valentini2017noisy}) to compare their models. 
%
%

Here, we also use the widely accepted VCTK evaluation dataset ($X$) as a clean reference.
%
We use the HiFi-GAN~\cite{su2020hifi} architecture to train enhancement models on two datasets: $M_1$ trained on DAPS~\cite{mysore2014can} and $M_2$ trained on VCTK~\cite{valentini2017noisy}.
Recordings from the VCTK set ($X$) are passed through the two models to produce $Y_1 = M_1(X)$ and $Y_2 = M_2(X)$.
We then compare $Y_1$ and $Y_2$ across various objective measures to assess performance. Finally, we do an MOS listening study to show how well different objective metrics correlate with subjective ratings. The results are summarized in Section~\ref{sec4.2}.

\input{LaTeX/tables/scenario3}
\subsection{Matching datasets acoustically}
\label{subsec2.3}


\fig{scenario3} illustrates a scenario wherein similarity metrics over-emphasize imperceptible differences such as frequency energy, inaudible noise and phase differences. To obtain parallel recordings coming from two different distributions, we start with the recordings $Y_1$ and $Y_2$ from \sect{subsec2.2}.
%
%
Next we process the output samples generated by DAPS model $Y_1$ so as to match the acoustics of the VCTK model $Y_2$, until they are almost perceptually indistinguishable, as follows.
%
%


%
%

\input{LaTeX/tables/differences_scenario2_new}
We first apply \emph{low-} and \emph{high-pass} filtering to remove unwanted frequency components, and prevent aliasing. Next we perform various steps, in order, based on the observation that $Y_1$ and $Y_2$ have different acoustics and background noise:
(1)~\emph{Equalization (EQ) matching} following Germain et al.~\cite{germain2016equalization} equalizes timbral content and background noise between audio recording environments. 
(2)~\emph{Breath removal} reduces perceptual differences by eliminating the minor breath sounds present in VCTK.
(3a)~\emph{Per-frequency energy normalization}~\cite{pandey2020cross} equalizes the energies at corresponding frequencies across samples. Because changing frequency energy may produce a ``frequency leak'' artifact if phase remains unchanged, we also experimented using 0, 200 and 1000 iterations of Griffin-Lim to refine the phase. 
(3b)~\emph{Per-frequency energy normalization} with ground-truth \emph{clean phase}. 
The results are summarized in Section~\ref{sec4.3}.

%
%

%% file: LaTeX/tables/scenario2.tex
\begin{figure}[t!]
\vspace{-1.2\baselineskip}
\centering
\setlength{\tabcolsep}{4pt}
\includegraphics[width=\columnwidth]{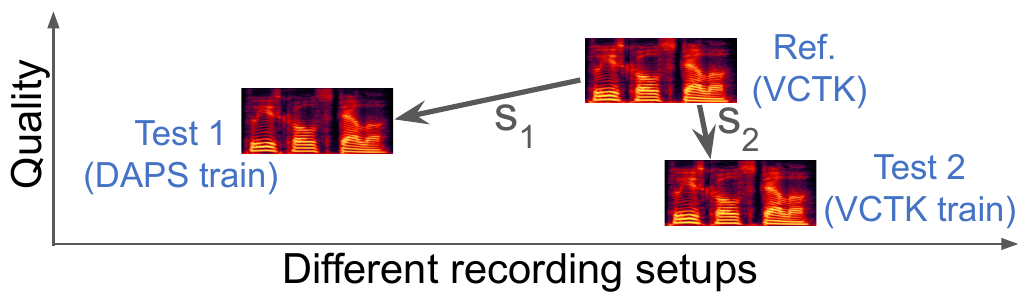}
\vspace{-0.25in}

\caption{\textbf{Scenario 2}: Test 1 and Test 2 are two recordings from models trained on two difference datasets (DAPS~\cite{mysore2014can} and VCTK~\cite{valentini2017noisy} respectively). Test 1 has a higher quality than the Test 2 recording as confirmed by listening tests. When compared across a reference that lies acoustically closer to Test 2 ($s_2<s_1$), we see that it may be more similar to Test 2, even though Test 1 has higher quality.}
\vspace{-0.10in}
\label{scenario2}
\end{figure}

%% file: LaTeX/tables/dds_datset_environment.tex
\begin{table*}[t!]
\vspace{-0\baselineskip}
\centering
\renewcommand{\arraystretch}{1.8}
\setlength{\tabcolsep}{3pt}
\resizebox{\textwidth}{!}{
 \begin{tabular}{$l ^c ^c ^c ^c ^c ^c ^c : ^c ^c ^c ^c : ^c | ^c ^c ^c ^c ^c ^c ^c : ^c ^c ^c ^c : ^c ^c ^c}
 \toprule
 \rowstyle{\largefontsize}
  \multirow{3}{*}{\bf \largefontsize Type} & \multicolumn{12}{c}{\bf \largefontsizelower DAPS~\cite{mysore2014can}} & \multicolumn{12}{c}{\bf \largefontsizelower VCTK~\cite{valentini2017noisy}} \\
 \cmidrule(lr){2-13} \cmidrule(lr){14-25} 
 \rowstyle{\largefontsizelower}
 &\bf PESQ &\bf VISQOL &\bf DPAM &\bf CDPAM &\bf ~~~~L1~~~~ &\bf ~~~~L2~~~~ &\bf M.STFT & \bf NISQA & \bf SQAPP  & \bf DNSMOS & \bf NORESQA & \bf MOS & \bf PESQ &\bf VISQOL &\bf DPAM &\bf CDPAM &\bf ~~~~L1~~~~ &\bf ~~L2~~ &\bf M.STFT & \bf NISQA & \bf SQAPP  & \bf DNSMOS & \bf NORESQA & \bf MOS\\
 \rowstyle{\largefontsize}
 & \bf $\uparrow$ &\bf $\uparrow$ &\bf $\downarrow$ &\bf $\downarrow$ &\bf $\downarrow$ &\bf $\downarrow$ &\bf $\downarrow$ & \bf $\uparrow$ & \bf $\uparrow$  & \bf $\uparrow$ & \bf $\downarrow$ & \bf $\uparrow$ & \bf $\uparrow$ &\bf $\uparrow$ &\bf $\downarrow$ &\bf $\downarrow$ &\bf $\downarrow$ &\bf $\downarrow$ &\bf $\downarrow$ & \bf $\uparrow$ & \bf $\uparrow$  & \bf $\uparrow$ & \bf $\downarrow$ & \bf $\uparrow$\\
 
 \cmidrule(lr){1-25}
 \rowstyle{\largefontsize}
  {\bf Clean}
 & - & - & -  & - & - & - & - & 4.68 & 3.45 & 3.85  & 9.56 & 4.48 &   & - & - & -  & - & - & - & 4.14 & 3.37 & 3.62 & 9.97 & 4.18 \\
 \rowstyle{\largefontsize}
 {\bf Confroom1}
 & 1.55 & 2.37 & 2.80 & 0.30 & 2.65 & 30.70 & 0.19 & 2.89 & 3.08 & 3.46 & 12.02 & 2.90 &  1.70 & 2.15 & 2.80 & 0.32 & 2.04 & 22.84 & 0.17  & 2.81 & 3.03 & 3.50 & 13.24 & 2.77\\
 \rowstyle{\largefontsize}
 {\bf Confroom2}
 & 1.33 & 2.20 & 2.79 & 0.34 & 2.76 & 31.37 & 0.20 & 2.38 & 2.773 & 3.17 & 13.02  & 2.39 & 1.48 & 2.04 & 2.75 & 0.38 & 2.12 & 23.31 & 0.19 & 2.37 & 2.75 & 2.93 & 14.35 & 2.29\\
 \rowstyle{\largefontsize}
 {\bf Office1}
 & 1.80 & 2.42 & 2.73 & 0.29 & 2.44 & 27.86 & 0.19 & 3.01 & 3.10 & 3.52 & 11.01 & 2.99 & 1.94 & 2.14 & 2.69 & 0.34 & 1.88 & 20.90 & 0.17 & 2.81 & 3.00 & 3.37 & 11.82 & 2.79\\
 \rowstyle{\largefontsize}
 {\bf Office2}
 & 1.57 & 2.38 & 2.77 & 0.32 & 2.52 & 28.96 & 0.19 & 2.71 & 3.04 & 3.42 & 11.52 & 2.63 & 1.70 & 2.12 & 2.75 & 0.37 & 1.95 & 21.74 & 0.18 & 2.64 & 2.97 & 3.32 & 12.49 & 2.60\\
 \rowstyle{\largefontsize}
 {\bf  Studio1}
 & 1.59 & 2.35 & 2.78 & 0.32 & 2.62 & 29.40 & 0.19 & 2.70 & 2.94 & 2.95 & 12.07 & 2.63 & 1.72 & 2.04 & 2.74 & 0.37 & 2.06 & 22.24 & 0.18 & 2.65 & 2.84 & 3.16 & 13.51 & 2.53\\
 \rowstyle{\largefontsize}
 {\bf Studio2}
 & 2.03 & 2.60 & 2.76 & 0.27 & 2.40 & 27.25 & 0.18 & 3.48 & 3.20 & 3.65 & 10.91 & 3.20 & 2.06 & 2.27 & 2.67 & 0.34 & 1.83 & 20.05 & 0.17 & 3.19 & 3.07 & 3.50 & 11.39 & 2.96\\
 \rowstyle{\largefontsize}
 {\bf  Studio3}
 & 2.03 & 2.54 & 2.83 & 0.29 & 2.53 & 30.50 & 0.17 & 3.58 & 3.18 & 3.61 & 10.86 & 3.28 & 2.03 & 2.23 & 2.78 & 0.33 & 1.96 & 22.86 & 0.16 & 3.29 & 3.03 & 3.41 & 11.16 & 3.03\\
 \rowstyle{\largefontsize}
 {\bf Waitingroom1}
 & 2.11 & 2.55 & 2.75 & 0.27 & 2.39 & 27.80 & 0.17 & 3.46 & 3.23 & 3.55 & 10.81 & 3.42 & 2.17 & 2.23 & 2.69 & 0.31 & 1.83 & 20.05 & 0.16 & 3.21 & 3.07 & 3.52 & 11.21 & 3.15\\
 \rowstyle{\largefontsize}
 {\bf Livingroom1}
 & 1.56 & 2.36 & 2.87 & 0.30 & 2.67 & 32.44 & 0.19 & 2.82 & 3.14 & 3.55 & 11.28 & 2.98 & 1.71 & 2.14 & 2.88 & 0.35 & 2.05 & 24.05 & 0.17 & 2.75 & 3.01 & 3.48 & 12.16 & 2.78\\
 \bottomrule
\end{tabular}
}
\caption{\textbf{DDS Dataset}: Performance of similarity, no-reference metrics and subjective MOS scores ($\pm$0.03) for recordings from the DAPS~\cite{mysore2014can} and VCTK~\cite{valentini2017noisy} datasets, re-recorded across various environments (DDS dataset)~\cite{li2021dds}). We see that both \textit{studio2} and \textit{studio3} show close MOS scores even though recorded under different environments. We also observe that the similarity metrics rate VCTK recordings as higher quality than the DAPS recordings, even though no-reference metrics and subjective ratings suggest otherwise.}


\vspace{-0.3in}
\label{tab:dds_environments}
\end{table*}

%% file: LaTeX/tables/scenario3.tex
\begin{figure}[t!]
\vspace{-1.2\baselineskip}
\centering
\setlength{\tabcolsep}{4pt}
\includegraphics[width=\columnwidth]{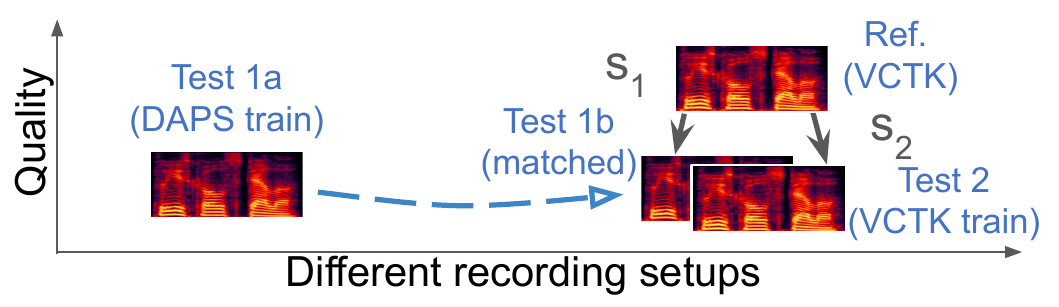}
\vspace{-0.25in}
\caption{\textbf{Scenario 3}: Test 1a and Test 2 are two recordings from models trained on different datasets (DAPS~\cite{mysore2014can} and VCTK~\cite{valentini2017noisy} respectively). Test 1b is created by acoustically matching recordings from Test 1a to Test 2. Even though recordings from Tests 1b and 2 sounds perceptually indistinguishable, they may have different similarities ($s_1$ and $s_2$) with a reference, which suggests these metrics emphasize imperceptible differences.}

%
\vspace{-0.1in}
\label{scenario3}
\end{figure}

%% file: LaTeX/tables/differences_scenario2_new.tex
\begin{table}[b!]
\vspace{-1\baselineskip}
\centering
\renewcommand{\arraystretch}{1.8}
\setlength{\tabcolsep}{3pt}
\resizebox{\columnwidth}{!}{
    \begin{tabular}{l ^c ^c ^c ^c ^c ^c ^c : ^c ^c ^c ^c}

 \toprule
 \multirow{1}{*}{\bf Type} & \bf PESQ & \bf VISQOL &\bf DPAM & \bf CDPAM &\bf L1 & \bf \largefontsize L2 & \bf \largefontsize M.STFT & \bf SQAPP & \bf NISQA  & \bf DNSMOS & \bf NORESQA \\
 
 
 & \bf $\uparrow$ &\bf $\uparrow$ &\bf $\downarrow$ &\bf $\downarrow$ &\bf $\downarrow$ &\bf $\downarrow$ &\bf $\downarrow$ & \bf $\uparrow$ & \bf $\uparrow$  & \bf $\uparrow$ & \bf $\downarrow$ \\
 
 \cmidrule(lr){1-12}
\rowstyle{\largefontsize}
 {\bf Studio2}
  & 1.81 & 3.32  &  1.79 &  0.13 & 0.75 & \largefontsize9.00 &  \largefontsize0.10 &  3.71  &  3.80 &  3.64 &  10.79  \\
\rowstyle{\largefontsize}
 {\bf Studio3}
  &  2.75 & 3.51 & 2.62 &  0.07 &  0.63 &  \largefontsize7.77 &  \largefontsize0.09 &  3.84 &  3.89 &  3.55 &  10.59  \\
\bottomrule
\end{tabular}
}

\caption{\textbf{Scenario 1}: Performance of similarity and no-reference metrics when reference recordings from \emph{studio2} and \emph{studio3}, and test recordings from \emph{confroom1} are selected. We see that similarity metrics show different similarities, even though no-reference metrics (and subjective ratings - Table~\ref{tab:dds_environments}) suggest the two references are of equal quality.}

\vspace{-0.30in}
\label{low_quality_sec4.2}
\end{table}

%% file: LaTeX/texts/04-results.tex
\section{Results}

\subsection{Two reference with one test recording}
\label{sec4.1}
Refer to Tables~\ref{tab:dds_environments} and~\ref{low_quality_sec4.2} for details. First, we see whether recordings from \textit{studio2} and \textit{studio3} are rated as equally good, or not. Table~\ref{tab:dds_environments} shows the scores~\footnote{The DDS Dataset~\cite{li2021dds} reports \textit{narrowband} \PESQ\, whereas Table~\ref{tab:dds_environments} reports \textit{wideband} \PESQ\ to focus on quality.} of a subjective listening study conducted on Amazon Mechanical Turk (AMT), where each subject is asked to rate the sound quality of an audio snippet on a scale of 1 to 5, with 1=Bad, 5=Excellent. Overall, we collect 1046 ratings per condition from 244 unique native english speakers. The workers have to pass a hearing test where they are asked to identify a word heard in a long sentence. This removes participants that either do not understand English, were not paying attention, or had a hearing impairment.
We observe that the MOS ratings for \textit{studio2} and \textit{studio3} are very close. Moreover, all considered objective metrics (including similarity and no-reference metrics) also show similar ratings. For this scenario, this suggests that: (i)~the recordings have equal quality; and (ii)~similarity metrics act as a good proxy for subjective ratings when compared within each individual dataset (DAPS and VCTK, respectively). 



Intuitively, reference recordings that have equal quality should not change the outcome of a quality metric.
However, it is not the case. Refer to Table~\ref{low_quality_sec4.2}. It reports all objective measures between clean recordings selected from \textit{studio2} or \textit{studio3}, and test recordings from \textit{confroom1}. Majority of similarity metrics show large (up to 85\%) differences, and fail to provide equal similarity ratings even when the reference recordings are judged to be equal quality. However, no-reference metric ratings are very close (up to 3.5\% difference), and reflect the MOS ratings well. Note that we consider a specific device (\emph{Marantz}) and position (\emph{ch1}) since we found certain other recordings corrupted.

\subsection{Single reference with two test recordings}
\label{sec4.2}
Refer to Table~\ref{table_similarity_comparison}. First, we see that various similarity metrics (including conventional - L1, L2, specific-feature-based - multi-resolution STFT and ML driven - \PESQ, \VISQOL, \DPAM\ and \CDPAM) rate the VCTK trained model higher than the DAPS trained model when evaluated on the VCTK evaluation dataset. 
However, when we evaluate with no-reference measures, we see a reversed trend.
Next, to see which measure reflects human-perceived quality, we conducted a subjective listening study on AMT, where each subject is asked to rate the sound quality of an audio snippet on a scale of 1 to 5, overall collecting around 1974 ratings per condition from 202 unique english speakers. Similar to previous experiment, the workers were screened based on being able to identify a word heard in a long sentence.

In Table~\ref{table_similarity_comparison}, we see that no-reference metrics and subjective listening ratings follow the same trend as human rating. For this scenario, this shows that: (i)~similarity measures do not reflect subjective ratings; and (ii)~no-reference metrics reflects subjective quality well.
%
We theorize similarity metrics are over-sensitive to acoustic difference between the two datasets. 
The only way to get rid of this discrepancy is to train and test models ``only'' on the same dataset(s) (to prevent a mismatch in acoustic settings). It is also worth noting that using a higher-quality dataset - DAPS, as compared to VCTK (see Table~\ref{tab:dds_environments} - row1) for training an enhancement model leads to a better model with higher audio quality (Table~\ref{table_similarity_comparison} last row). However, if using similarity metric as proxy for quality, higher-quality models may be labelled as lower quality.

\input{LaTeX/tables/modelDAPSvsmodelVCTK}

\subsection{Matching datasets acoustically}
\label{sec4.3}

Refer to Table~\ref{pipeline}. Recall that we developed a set of pre-processing stages to acoustically match the recordings from the DAPS trained model to the VCTK trained model used in \sect{sec4.2}. For all stages, we report all objective metrics, and subjective listener ratings. The subjective listening tests are conducted on AMT, where a subject is asked to rate the quality of an audio snippet on a scale of 1 to 5, collecting around 1974 ratings per condition from 371 unique english speakers. Similar to previous experiments, the workers were screened based on being able to identify a word heard in a long sentence.

We see that all objective similarity metrics appear to be converging to the target VCTK Model scores (row1).
However, this trend breaks down when as we use griffin-lim to estimate the phase.
Comparing scores across the \textit{Breath rem.} and \textit{itr0} stages, we observe a sharp ($\approx$ 15\%) decrease in \PESQ\ scores even though MOS ratings are consistent (within 2 standard deviations, differences not statistically significant). Moreover, comparing the scores across two stages - \emph{itr0}, and \emph{original phase}, we note a significant difference in scores across all similarity metrics (\eg\ 36\% in PESQ). However, we do not see a similar difference in MOS ratings for these stages which suggests that similarity metrics (including advanced metrics like \DPAM\ and \CDPAM) emphasize imperceptible differences (\eg\ small phase differences) and fail to reflect absolute quality even when the two recordings are almost the same. Furthermore, we see that no-reference metrics do not show the same shortcomings, but they are insensitive to reduced quality in \emph{itr200} and \emph{itr1000} possibly caused by artifacts introduced from Griffin-Lim.
\input{LaTeX/tables/misc_spectrograms}

\input{LaTeX/tables/scenario3_pipeline}

%


\input{LaTeX/tables/different_types_PESQ}
\subsection{Miscellaneous Observations}
\label{misc_observations}
We show several other scenarios where similarity metrics (esp. \PESQ) work well, and others that highlight their drawbacks.

\noindent {\bf Reversed correlation between PESQ and MOS}
\label{DDS_dataset}
Looking at Table~\ref{tab:dds_environments} specific to \PESQ\ (and other metrics like \DPAM, L1, L2, Multi-res. STFT), we observe that the DAPS quality still remains lower than the corresponding VCTK quality for each environment condition, even if subjective listening suggests a reverse trend. This suggests that \PESQ\ (and other metrics) cannot be used to compare absolute qualities across two different datasets (\eg\ a \PESQ\ score of 1.55 in \textit{confroom1} DAPS v/s 1.70 VCTK, even when no-reference metrics and MOS ratings suggests otherwise).

\noindent {\bf High-pass filtering}
Refer to Figure~\ref{model}. We apply a high pass filter (butterworth filter, order 5) of increasing cutoff frequencies (upto 250Hz) on a test recording, and compare it to a given reference audio recording. We observe that even though the spectrum looks very different, \PESQ\ (and other metrics like \DPAM, \CDPAM, L1, L2)
predict a higher rating with the high-pass filtered test recording as compared to the rating for the clean test recording.

\noindent {\bf Comparing cross-perturbations}
Refer to Figure~\ref{model_pesq_compare}. We apply different perturbations to the reference recording, and create two test samples: \emph{Audio1} contains background white noise, and \emph{Audio2} contains compression artifacts. Audio1 sounds closer to the reference according to human judgments. In contrast, similarity metrics find Audio2 closer to the reference (\eg\ PESQ score of 2.57 for Audio1 v/s 3.31 for Audio2, even when subjective listening suggests otherwise) suggesting that these metrics cannot reliably evaluate recordings across perturbations.

\noindent {\bf Effect of a noisy reference}:
 We apply band-limited white noise (between 0 and 3.5kHz) to the reference recording, and observe an increase in \PESQ\ score of upto 0.1 points (Figure~\ref{ref_noise_added}). Furthermore, when we apply the same noise to both the reference and test recordings (Figure~\ref{ref_noise_added}), we observe an increase in \PESQ\ of upto 0.2 points. These observations suggest that \PESQ\ can match noise characteristics between signals to output a higher rating.

\noindent {\bf Variation with frame-wise alignment}
We found \PESQ, and other perceptual similarity metrics like \VISQOL, \DPAM\ and \CDPAM\ to be quite stable with small mis-alignment in the signal. In contrast, conventional similarity metrics like L1, L2 and multi-resolution STFT show high variance with small mis-alignments.








\input{LaTeX/tables/ref_noise_added_combined}




%% file: LaTeX/tables/modelDAPSvsmodelVCTK.tex
\begin{table}[b!]
\vspace{-1\baselineskip}
\renewcommand{\arraystretch}{1.8}
\setlength{\tabcolsep}{3pt}
\resizebox{\columnwidth}{!}{ 
 \begin{tabular}{$l ^c ^c ^c ^c}
 \toprule
 \rowstyle{\largefontsize}
 {\bf } &  {\bf DAPS Model} & {\bf VCTK Model} & {\bf Valset Noisy} & {\bf Valset Clean}
  \\ \toprule
  \rowstyle{\largefontsize}
  {\bf PESQ $\uparrow$ } 
 &  2.16 & 3.13 & 1.96 & -  \\
 \rowstyle{\largefontsize}
 {\bf VISQOL $\uparrow$} 
 \rowstyle{\largefontsize}
 &  3.60 & 4.18 & 3.81 & -  \\
 \rowstyle{\largefontsize}
 {\bf DPAM $\downarrow$} 
 &  2.77 & 1.35 & 1.71 & -  \\
 \rowstyle{\largefontsize}
 {\bf CDPAM $\downarrow$} 
 &  0.21 & 0.07 & 0.30 & -  \\
\cdashline{1-5}
\rowstyle{\largefontsize}
  {\bf L1 $\downarrow$} 
 &  2.24 & 0.30 & 0.92 & - \\
 \rowstyle{\largefontsize}
 {\bf L2 $\downarrow$} 
 &  20.42 & 2.14 & 5.90 & \bf -  \\
 \rowstyle{\largefontsize}
 {\bf Multi-res STFT $\downarrow$ }
 & 0.14 &  0.07 &  0.17 & - \\
 \cdashline{1-5}
 \rowstyle{\largefontsize}
  {\bf SQUAPP $\uparrow$} 
  \rowstyle{\largefontsize}
 &  4.06 & 3.76 & 2.73 & 3.85 \\
 \rowstyle{\largefontsize}
 {\bf NISQA $\uparrow$} 
 &  4.90 & 4.65 & 3.06 &  4.60 \\
 \rowstyle{\largefontsize}
 {\bf DNSMOS $\uparrow$}
 \rowstyle{\largefontsize}
 &  3.64 &  3.55 &  3.02 & 3.55 \\
 \rowstyle{\largefontsize}
 {\bf NORESQA $\downarrow$}
 \rowstyle{\largefontsize}
 & 9.57 & 10.53 & 12.88 & 9.44  \\
 \cdashline{1-5}
 \rowstyle{\largefontsize}
  {\bf MOS$\uparrow$} 
 & 4.10 & 4.02 & 2.48 & 4.29 \\
 \bottomrule
\end{tabular}
}

\caption{\textbf{Scenario 2}: Performance of similarity metrics, no-reference metrics and MOS ratings ($\pm$0.02) across speech enhancement (SE) models trained on two datasets (DAPS and VCTK), and evaluated on the VCTK evaluation set.} 
\vspace{-0.0in}
\label{table_similarity_comparison}
\end{table}

%% file: LaTeX/tables/misc_spectrograms.tex
\begin{figure}[t!]
\vspace{-1\baselineskip}
\centering
\setlength{\w}{0.33\columnwidth}
\setlength{\tabcolsep}{0pt}
\begin{tabular}{ccc}
\hspace{-0.2in}
\includegraphics[width=\w]{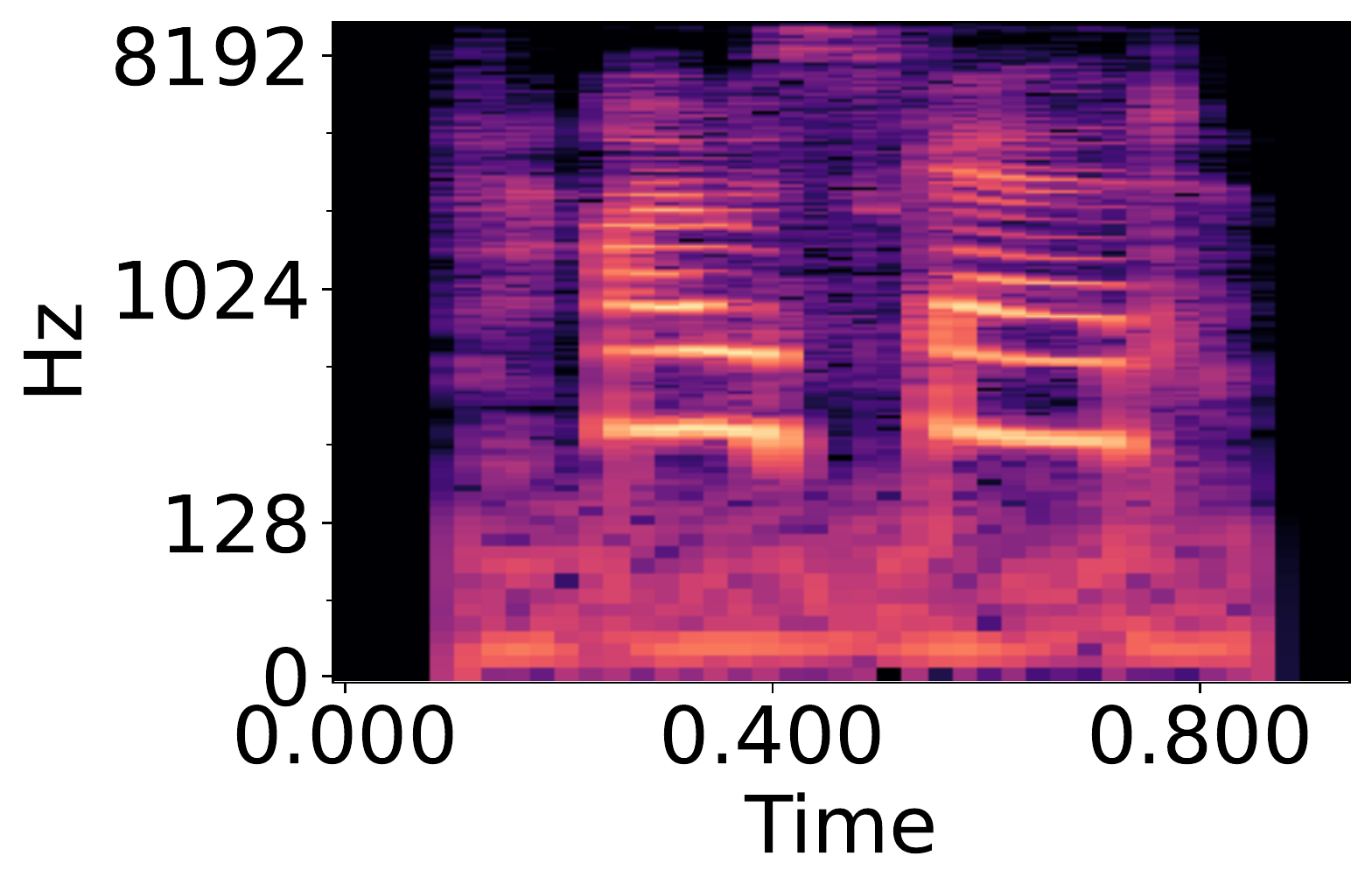} &
\hspace{-0.2in}
\includegraphics[width=\w]{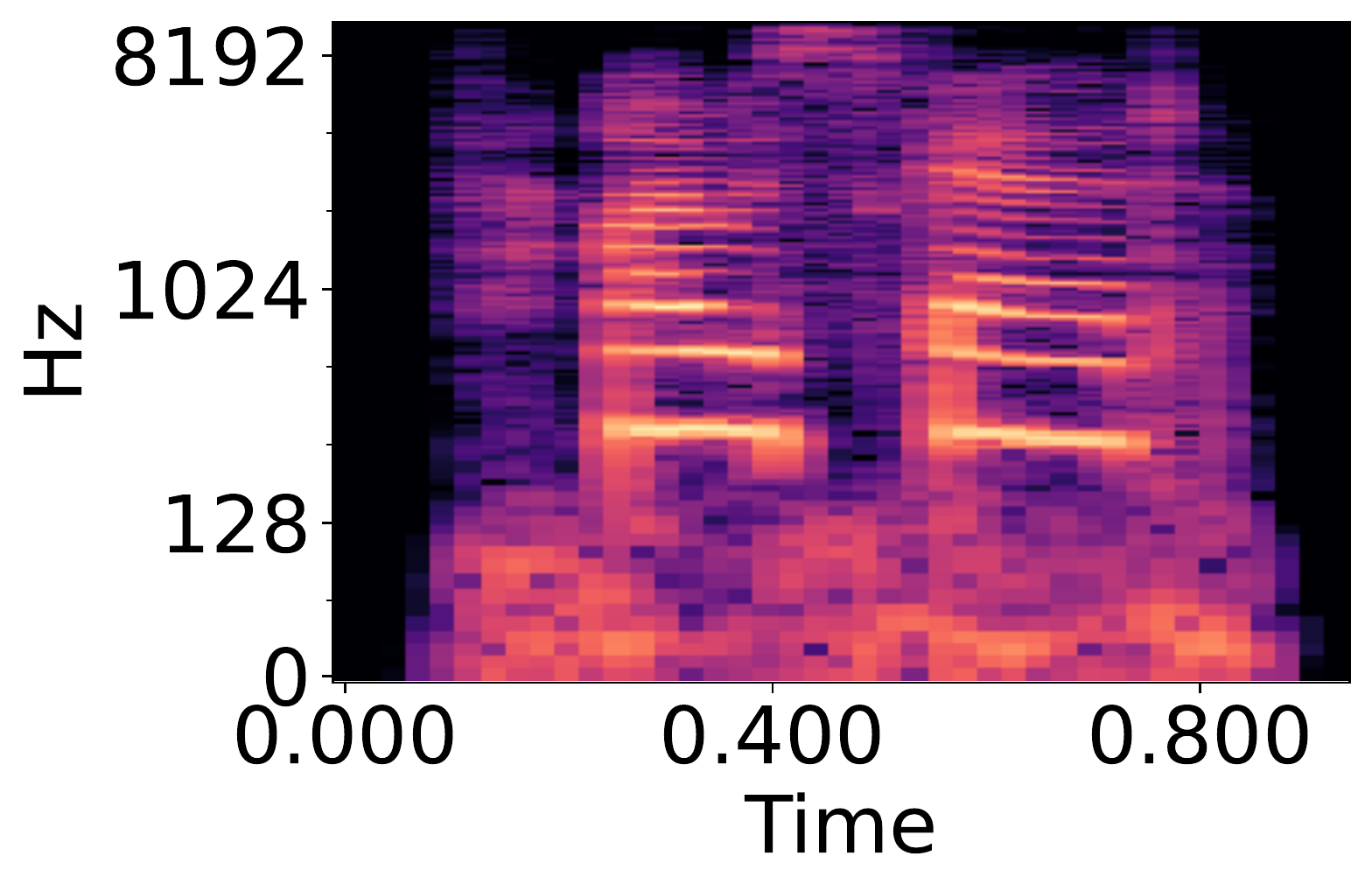} &
\hspace{-0.2in}
\includegraphics[width=\w]{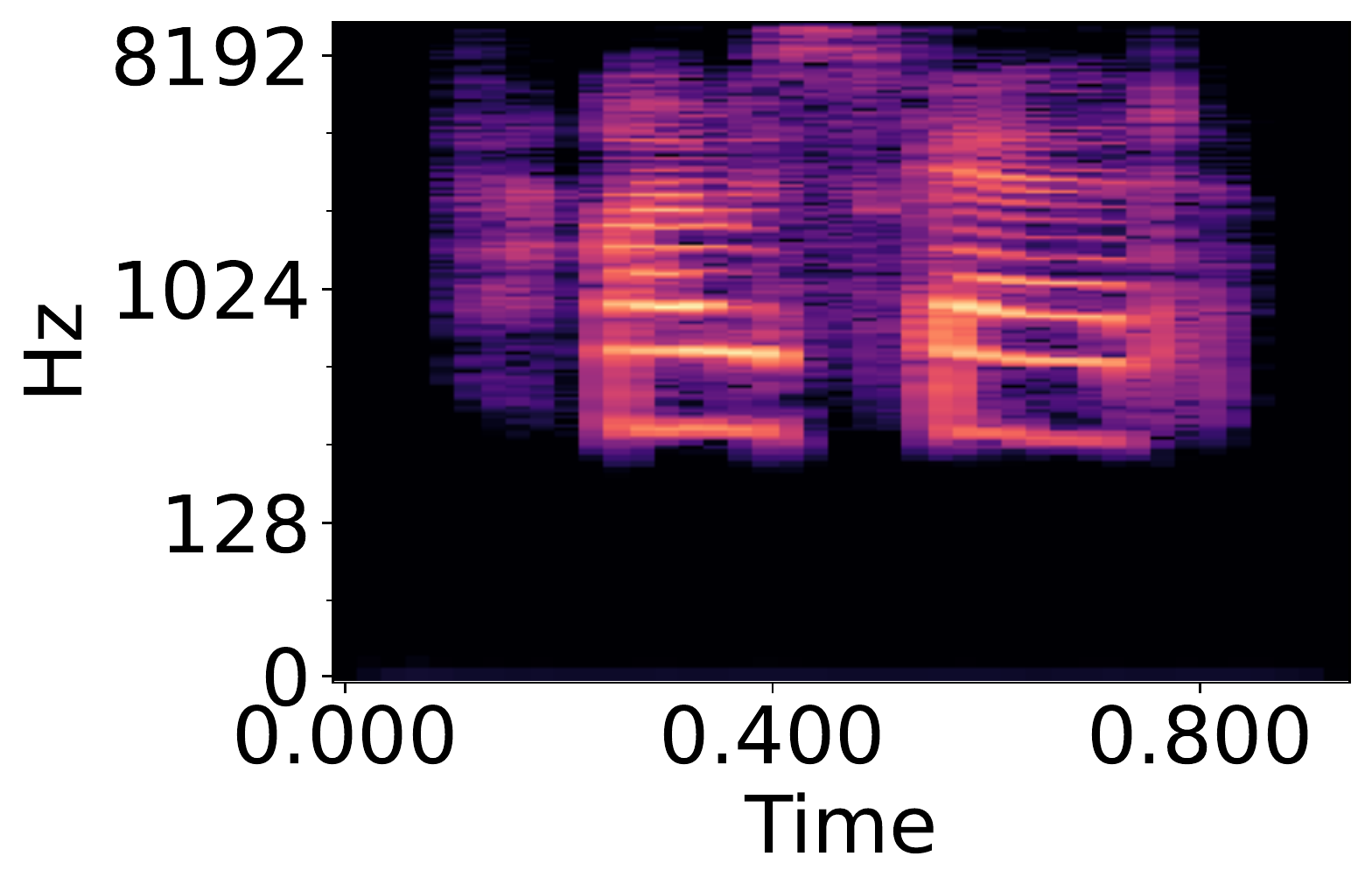} \\
\begin{minipage}{\w}\centering
\centering
{Clean reference}
\end{minipage} &
\begin{minipage}{\w}\centering
\centering
{Test recording} 
\end{minipage} &
\begin{minipage}{\w}\centering
\vspace{2pt}
{Test rec.+ high pass filter at 250Hz}
\end{minipage}
\end{tabular}
\vspace{-1.5ex}
\caption{\textbf{High-pass filter}: Reference, Test, and Test + high-pass. filtered recordings shown. \PESQ\ score increases from 2.54 to 2.84, even if visual and perceptual similarity suggests otherwise.}
\label{model}
\end{figure}

%% file: LaTeX/tables/scenario3_pipeline.tex
\begin{table}[t!]

\vspace{-0.25\baselineskip}
\centering
\renewcommand{\arraystretch}{1.8}
\setlength{\tabcolsep}{3pt}
\resizebox{\columnwidth}{!}{
 \begin{tabular}{$l ^c ^c ^c ^c ^c ^c ^c : ^c ^c ^c ^c : ^c ^c ^c ^c ^c ^c ^c ^c ^c ^c ^c}
  \toprule
  \multirow{1}{*}{\bf \largefontsize Type}
  \rowstyle{\largefontsizelower}
 &\bf PESQ &\bf VISQOL &\bf DPAM &\bf CDPAM &\bf L1 &\bf L2 &\bf M.STFT & \bf SQAPP & \bf NISQA  & \bf DNSMOS & \bf NORESQA  & \bf MOS \\
\rowstyle{\largefontsize}
 &\bf $\uparrow$ &\bf $\uparrow$ &\bf $\downarrow$ &\bf $\downarrow$ &\bf $\downarrow$ &\bf $\downarrow$ &\bf $\downarrow$ & \bf $\uparrow$ & \bf $\uparrow$  & \bf $\uparrow$ & \bf $\downarrow$  & \bf $\uparrow$ \\
 \cmidrule(lr){1-13}
 \rowstyle{\largefontsize}
 {\bf VCTK Model} 
 &  \largefontsize 3.13 & 4.18 & 1.35 & 0.07 & 0.30 & 2.14 & 0.07 & 3.76 & 4.65 & 3.55 & 10.53 & 4.02 \\
  \hline
  \rowstyle{\largefontsize}
{\bf DAPS Model} 
 & 2.16 & 3.60 & 2.77 & 0.21 & 2.24 & 20.42 & 0.14 & 4.06 & 4.90 & 3.64 & 9.57 & 4.10\\
 \rowstyle{\largefontsize}
 {\bf \hspace{1mm} EQ Match.} 
 & 2.32 & 3.68 & 2.73  & 0.22 & 2.00 & 25.99 & 0.13 & 3.96 & 4.89 & 3.61 & 10.32 & 4.11 \\
 \rowstyle{\largefontsize}
 {\bf \hspace{1mm} Breath rem.} 
 & 2.67 & 4.18 & 2.67 & 0.13 & 1.95 & 25.97 & 0.12 & 3.84 & 4.84 & 3.63 &  10.29 & 4.19\\
 \rowstyle{\largefontsize}
 {\bf \hspace{1mm} Energy norm.} \\
 \rowstyle{\largefontsize}
 {\bf \hspace{3mm} itr0} 
 & 2.32 & 4.17 & 2.68  & 0.14 & 1.92 & 24.85  & 0.13 & 3.87 & 4.77 & 3.56 & 10.09 & 4.14\\
 \rowstyle{\largefontsize}
 {\bf \hspace{3mm} itr200} 
 & 2.29 & 4.17 & 2.69  & 0.14 & 1.92 & 24.83 & 0.14 & 3.94 & 4.78 & 3.55 & 10.07 & 4.10 \\
 \rowstyle{\largefontsize}
 {\bf \hspace{3mm} itr1000}
 & 2.29 & 4.17 & 2.69  & 0.15 & 1.91 & 24.77 & 0.14 & 3.95 & 4.78 &  3.55 &  10.07 & 4.10\\
 \rowstyle{\largefontsize}
 {\bf \hspace{1mm} Orig. Phase}
 & 3.16 & 4.44 & 1.21 & 0.05 & 0.21 & 0.25 & 0.13 & 3.97 & 4.78 & 3.60 &  10.35 & 4.19\\
 \bottomrule
\end{tabular}
}
\vspace{0.05in}
\caption{\textbf{Scenario 3}: Objective measures and MOS ratings~($\pm$0.03) across pre-processing stages (\sect{subsec2.3}) when recordings from DAPS trained SE model are matched to the VCTK trained SE model.}
\vspace{-1.5\baselineskip}
\label{pipeline}
\end{table}

%% file: LaTeX/tables/different_types_PESQ.tex
\begin{figure}[b!]
\vspace{-0.75\baselineskip}
\centering
\setlength{\w}{0.33\columnwidth}
\setlength{\tabcolsep}{0pt}
\begin{tabular}{ccc}
\hspace{-0.2in}
\includegraphics[width=\w]{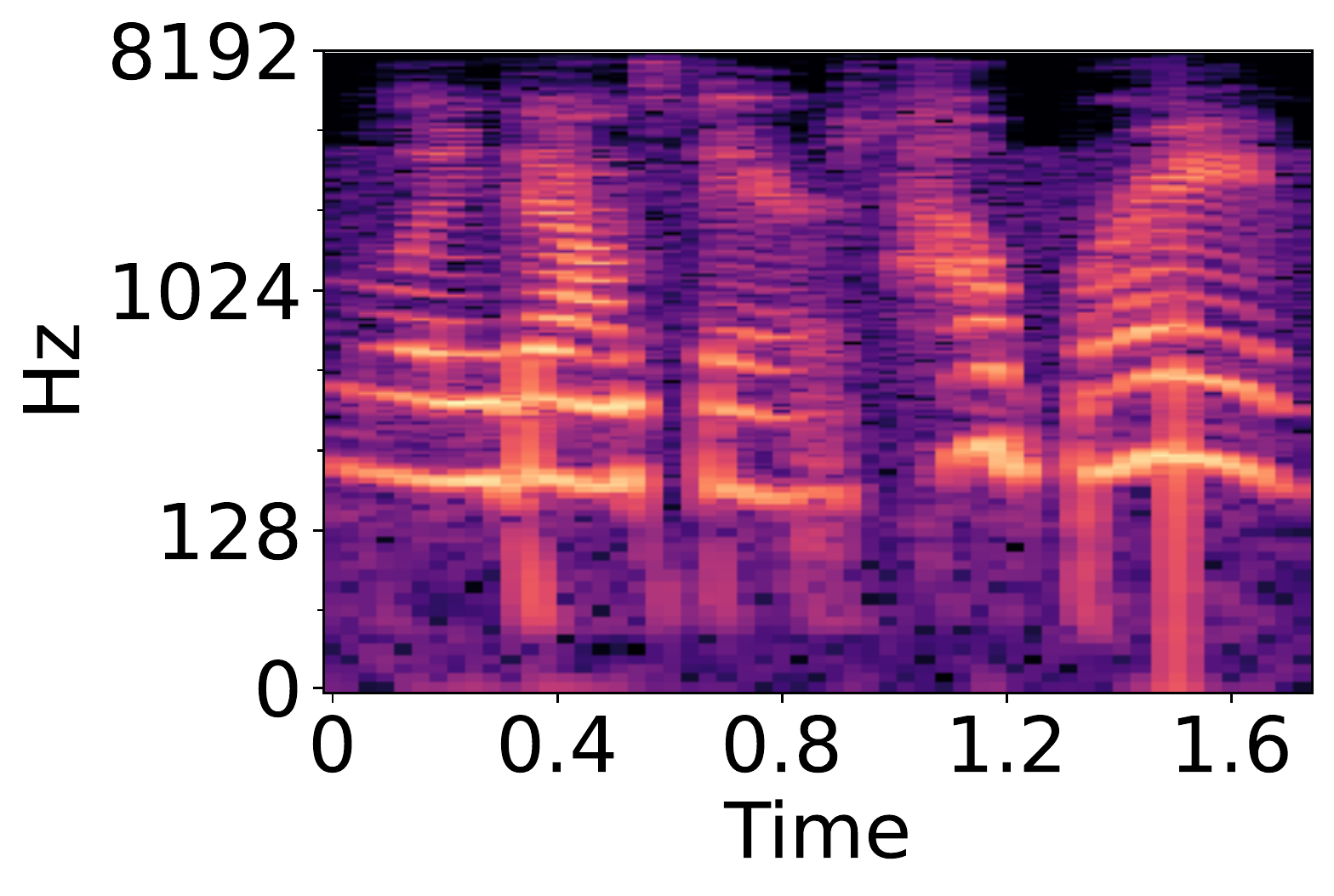} &
\hspace{-0.2in}
\includegraphics[width=\w]{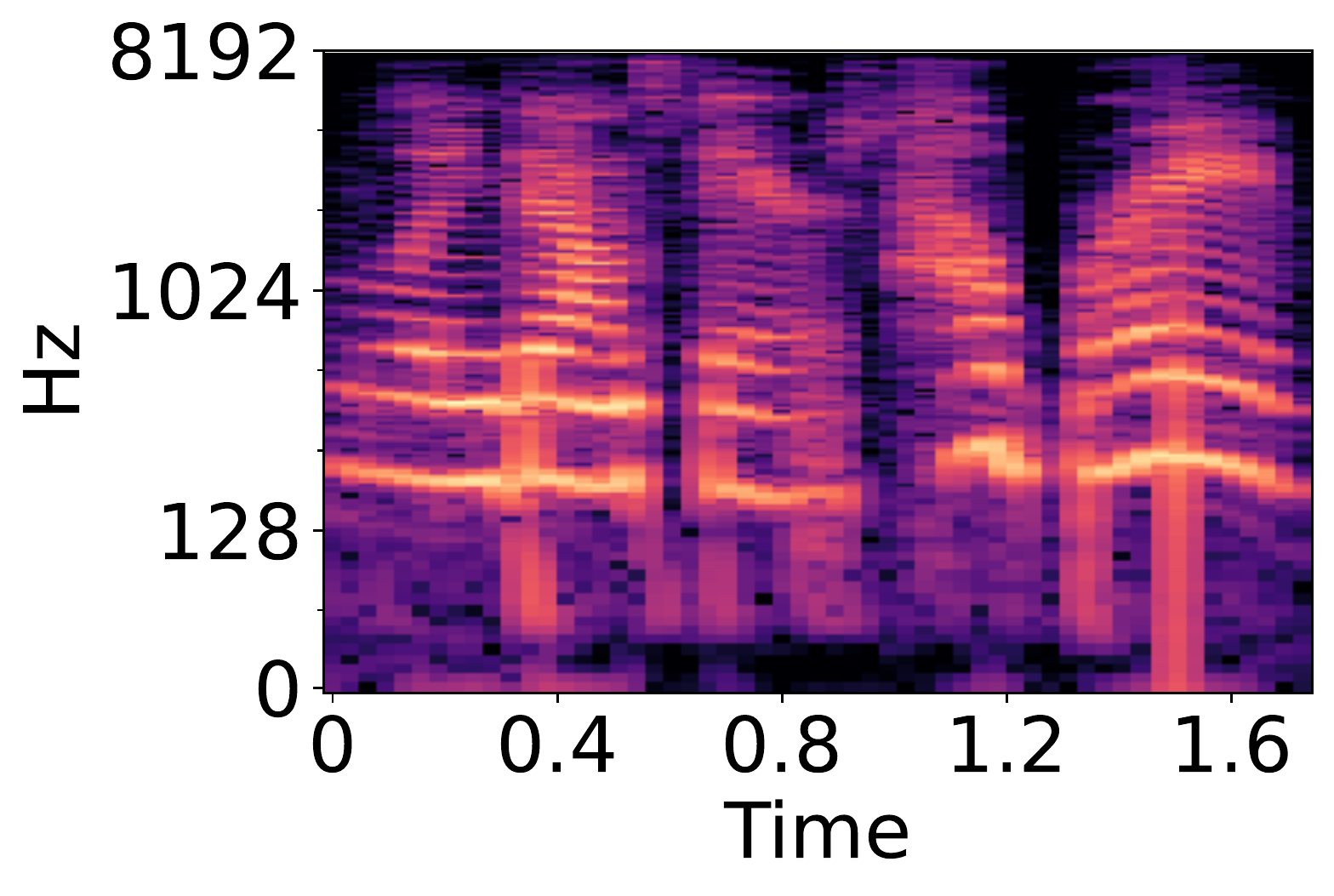} &
\hspace{-0.2in}
\includegraphics[width=\w]{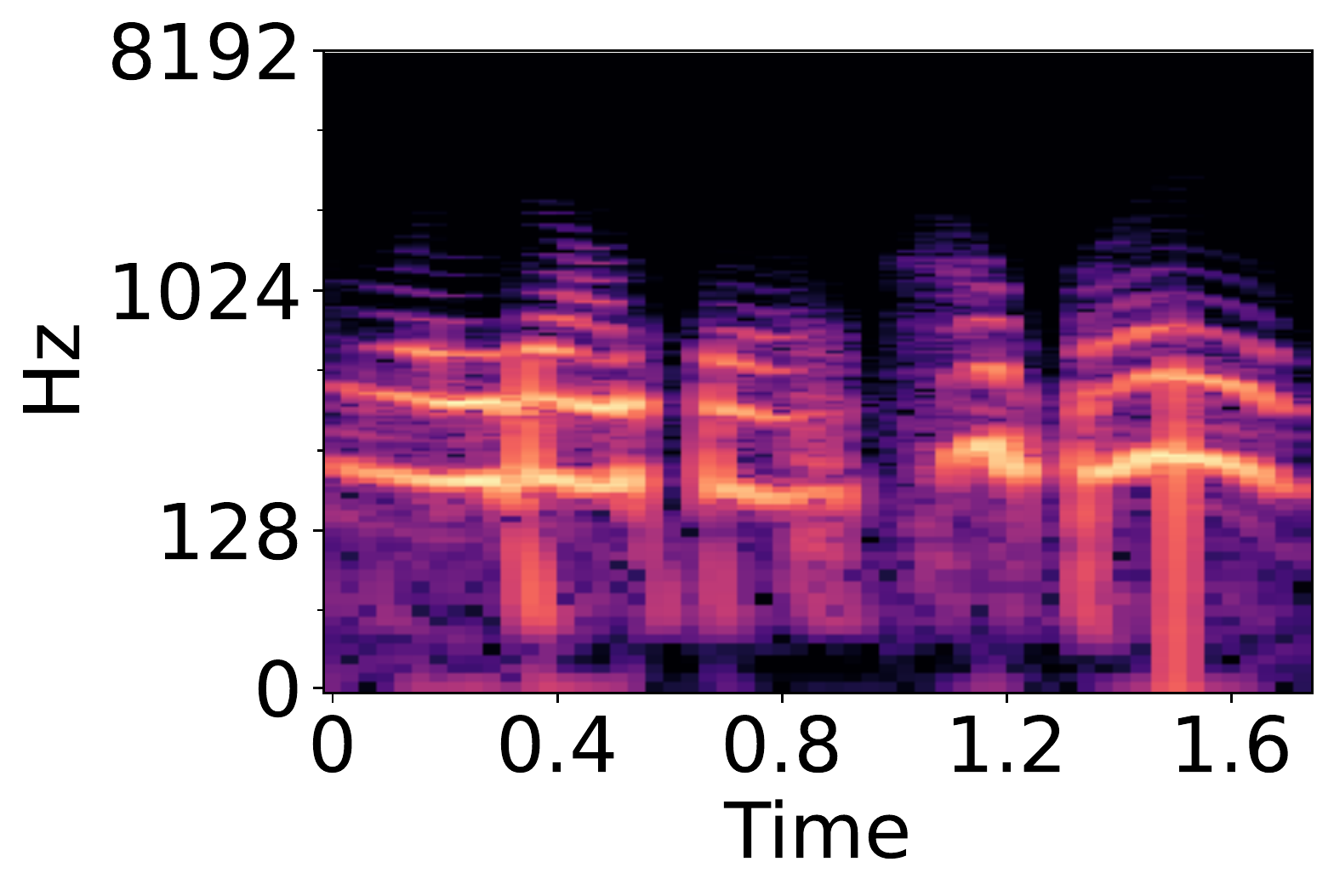} \\

\begin{minipage}{\w}\centering
\vspace{2pt}
\hspace{2mm}{\footnotesize {\bf Audio1} closer to Ref. acc. to {\bf humans}}

\end{minipage} &
\begin{minipage}{\w}\centering
\centering
\hspace{2mm}{\footnotesize {\bf Reference} speech recording. } 
\end{minipage} &
\begin{minipage}{\w}\centering
\vspace{2pt}

\hspace{2mm}{\footnotesize {\bf Audio2} closer to Ref. acc. to L1, L2,
\PESQ }

\end{minipage}
\end{tabular}
\vspace{-1.5ex}
\caption{\textbf{Cross-perturbation comparison}: Audio1 contains white noise, whereas Audio2 is band-limited: 
which one sounds ``closer'' to the reference recording? 
Similarity metrics struggle to perceptually compare artifacts of different types. Figure adapted from Manocha \etal~\cite{manocha2020differentiable}}
\label{model_pesq_compare}
\vspace{-1ex}
\end{figure}

%% file: LaTeX/tables/ref_noise_added_combined.tex
\begin{figure}[t!]
\vspace{-1.\baselineskip}
\centering
\setlength{\w}{0.33\columnwidth}
\setlength{\tabcolsep}{0pt}
\begin{tabular}{ccc}
\hspace{-0.2in}
\includegraphics[width=\w]{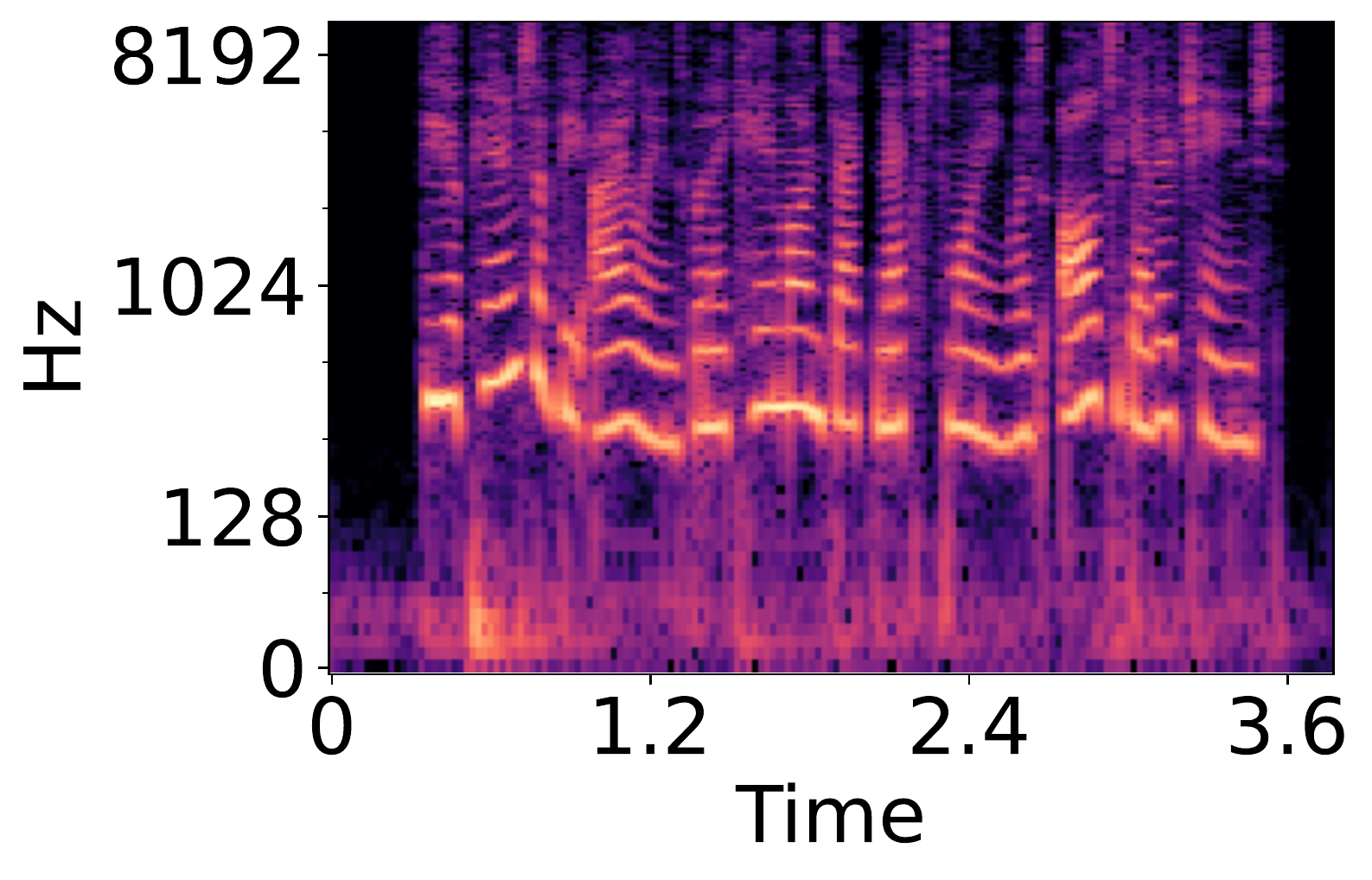}&
\hspace{-0.2in}
\includegraphics[width=\w]{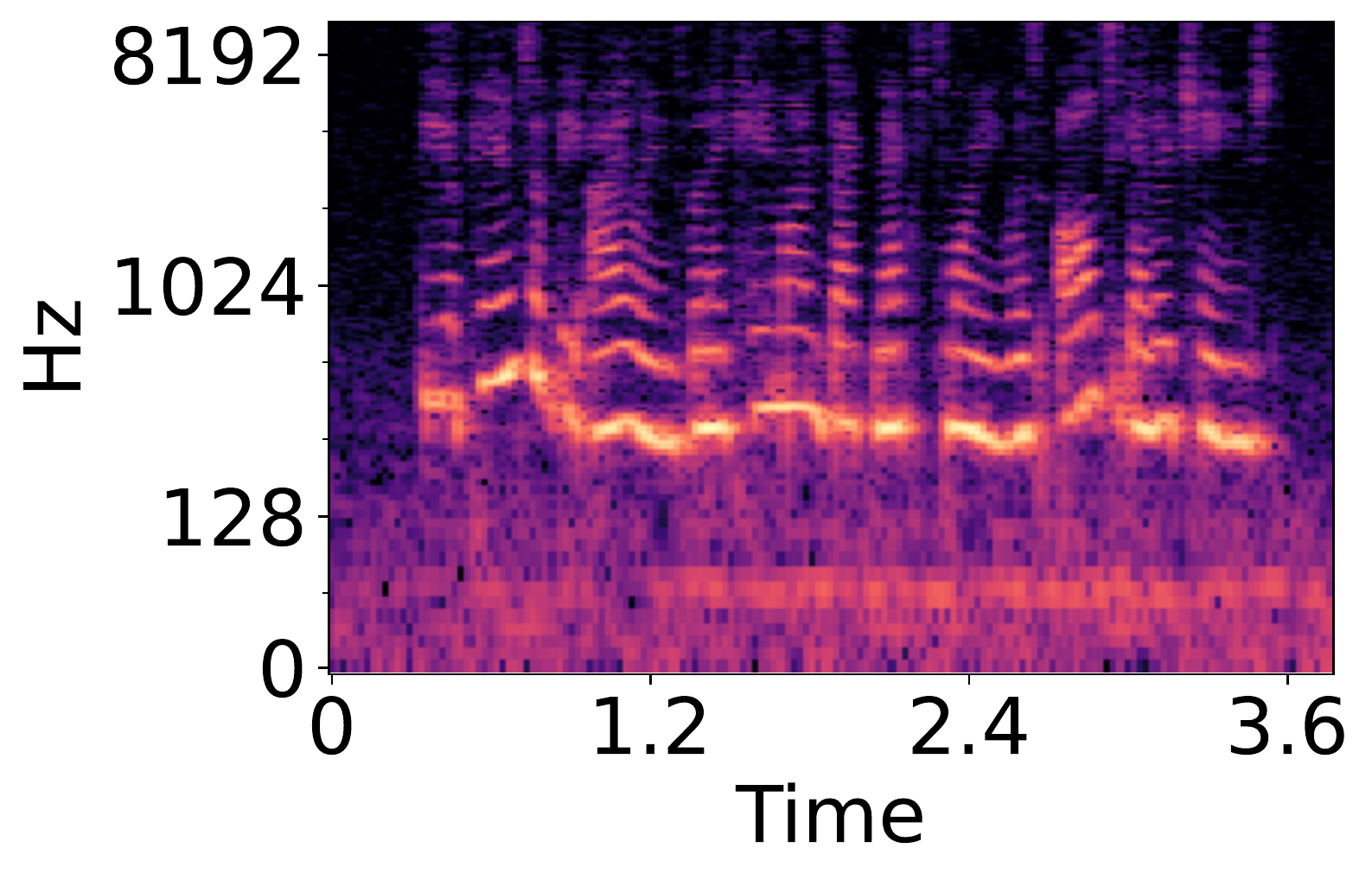} &
\hspace{-0.2in}
\includegraphics[width=\w]{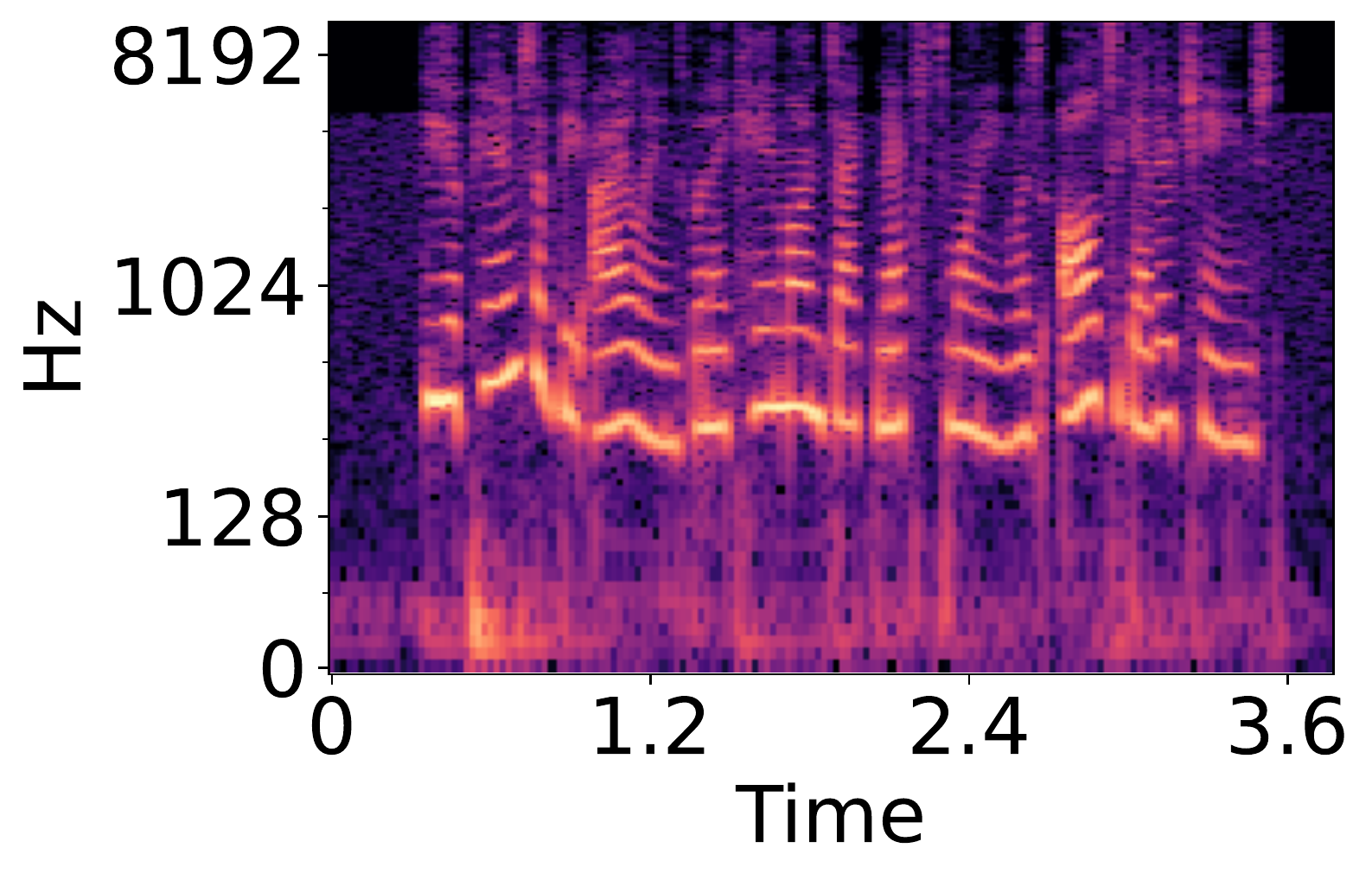} 
\\
\hspace{-0.2in}
\includegraphics[width=\w]{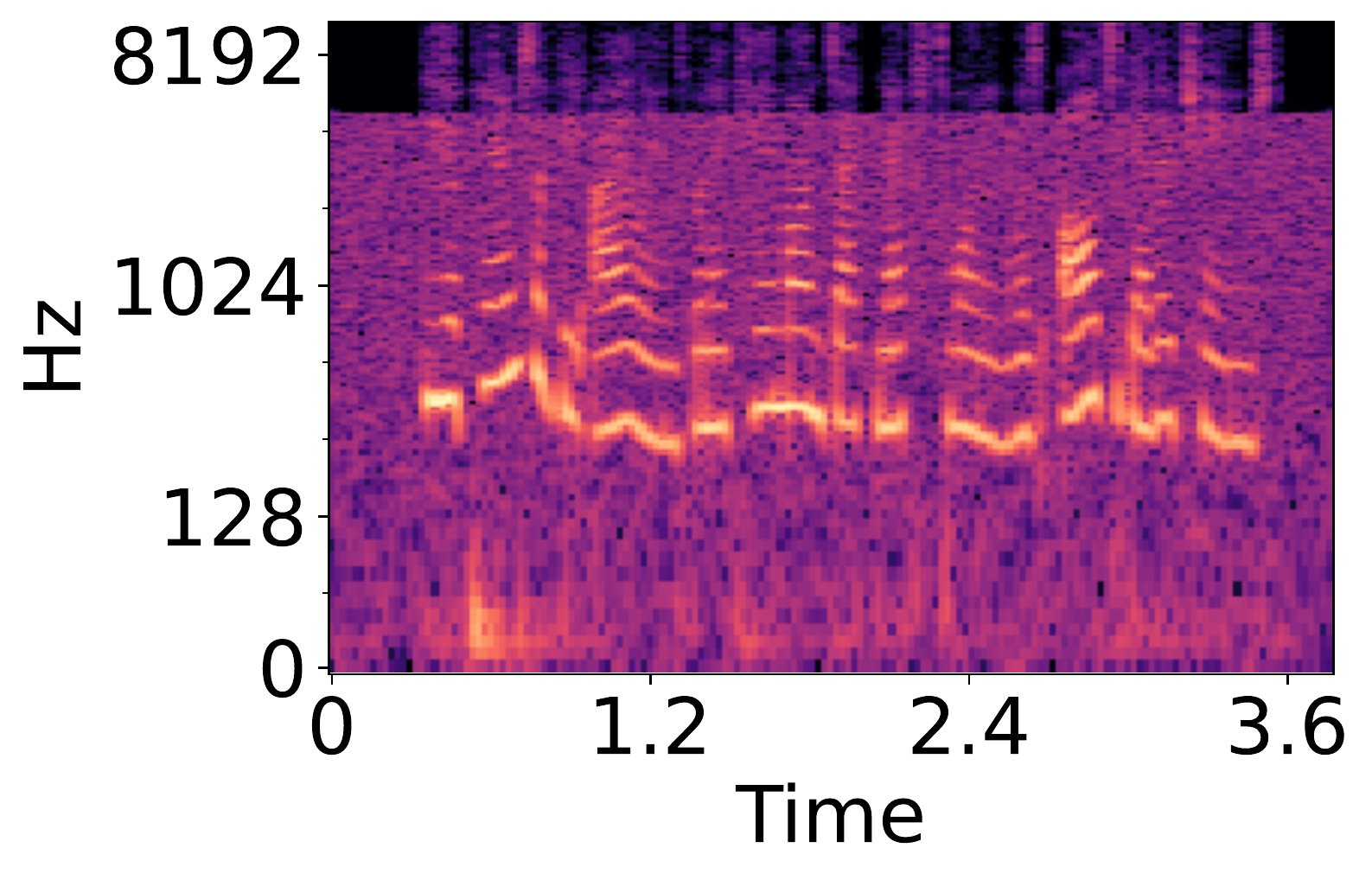} & 
\hspace{-0.2in}
\includegraphics[width=\w]{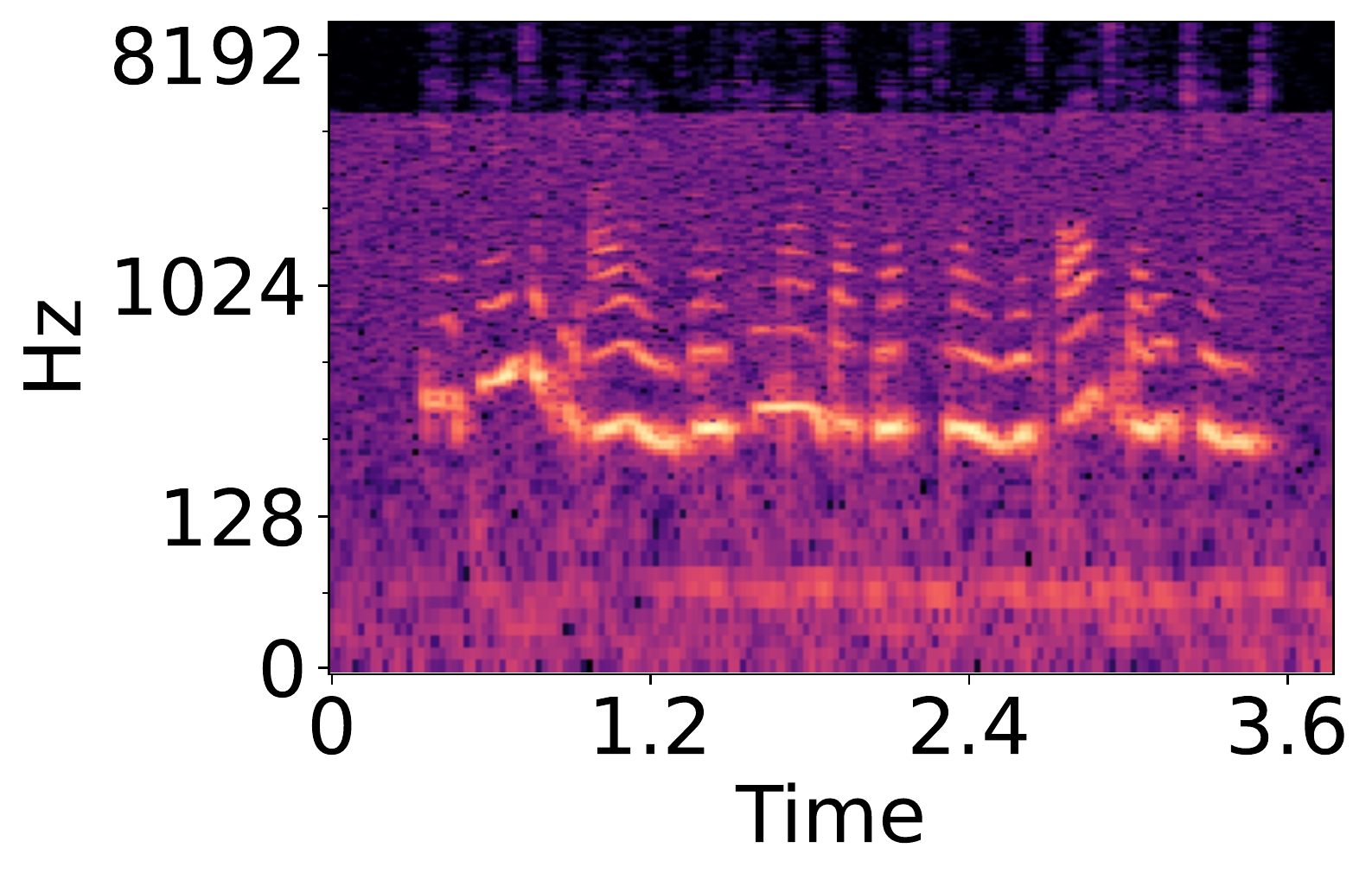}
&
\begin{minipage}{\w}\centering
{\vspace{-0.7in}Ref. rec. + white 
band-limited noise}
\end{minipage}

\\
\begin{minipage}{\w}\centering
\centering
{Reference rec.}
\end{minipage} &
\begin{minipage}{\w}\centering
\centering
{Test rec.} 
\end{minipage} &
\end{tabular}
\vspace{-1.5ex}

\caption{\textbf{Effect of Noise}: \emph{Top Row} (left to right): original reference and test signals - band-limited white noise added to reference recording - \PESQ\ score increases by up to 0.1 points; \emph{Bottom Row}: same band-limited white noise added to both reference and test signals. \PESQ\ score increases by up to 0.2 points.}
\label{ref_noise_added}
\vspace{-2.0ex}
\end{figure}

%% file: LaTeX/texts/06-conclusions.tex
\section{Conclusions and future work}

In this paper we discussed several scenarios where the conventional formulation of similarity metrics (\eg\ \PESQ) contrasts from predicting absolute subjective quality, and showed that various no-reference metrics are a good alternative in those settings. Overall, our findings suggest that similarity metrics (like \PESQ) are an unreliable proxy for audio quality, and should be used cautiously. In the future, we would like to exhaustively compare the performance of no-reference metrics with full-reference metrics, and see how well they correlate with human perception, especially on out-of-domain tasks.